\documentstyle[psfig]{mn}

\newif\ifAMStwofonts

\title[Spiral Structures in CV Accretion Discs]
      {On the observability of spiral structures in CV accretion discs}
\author[Steeghs and Stehle]{D.~Steeghs$^{1}$, R.~Stehle$^{2}$ \\
        $^{1}$Physics and Astronomy, University of St. Andrews, North Haugh,
        St. Andrews, KY16 9SS, U.K. \\
        $^{2}$University of Leicester, Astronomy Group, 
        Leicester LE1 7RH, U.K.}

\pagerange{\pageref{firstpage}--\pageref{lastpage}}
\pubyear{1999}
\date{Accepted 24 Feb 1999}

\begin{document}

\label{firstpage}

\maketitle

\begin{abstract}

We use the grid of hydrodynamic accretion  disc calculations of Stehle
(1999) to construct orbital  phase dependent emission line profiles of
thin    discs   carrying spiral   density    waves.  The observational
signatures of spiral  waves are explored  to establish the feasibility
of  detecting   spiral  waves in   cataclysmic   variable  discs using
prominent emission lines in the visible range of the spectrum.
For high Mach number accretion discs ($M = v_{\phi} / c_{\mathrm{s}}
\simeq$  15 -- 30), we find  that the spiral  shock arms are so tightly
wound that they leave few obvious  fingerprints in the emission lines.
Only a  minor  variation of the double  peak  separation in   the line
profile at a level of $\sim$8\% is produced.
For accretion discs in  outburst ($M \simeq  $ 5  -- 20)  however, the
lines  are dominated by  the emission from  an $m$=2 spiral pattern in
the disc.   We show that   reliable Doppler tomograms of  spiral shock
patterns  can be reconstructed  provided that a  signal to noise of at
least 15, a wavelength resolution of $\sim$ 80 km  s$^{-1}$ and a time
resolution of $\sim$ 50 spectra per binary orbit is achieved.
We confirm that the  observed spiral pattern in the  disc of IP Pegasi
can  be reproduced by  tidal density waves  in  the accretion disc and
demands the presence   of a large, hot  disc,  at least in   the early
outburst stages.

\end{abstract}


\section{Introduction}

Cataclysmic variables   (CVs)  harbour a  geometrically  thin  viscous
accretion  disc through which mass from  a late type companion star is
accreted  onto a white  dwarf.  Rapid brightness variations of several
magnitudes displayed by  the sub  class  of dwarf  novae indicate that
information is transported along the accretion disc  on a viscous time
scale of  a few days  (see Warner 1995 for an  extensive review).  The
dynamical time scale of  disc material on  the other hand, is given by
the local  Kepler  velocity and ranges  from   seconds near the  white
dwarf, to a considerable fraction of the orbital period (hours) at the
outer disc.  The effective viscous transport  displayed by these discs
demands the presence of a highly efficient  source of angular momentum
transport    within the  discs,    whose  physical  origin  is  poorly
understood.

The convenient   time scales in  CVs allow  us   to study the changing
properties of viscous accretion  discs in real-time.  As the accretion
disc is  viewed  from different angles  during  a binary  orbit, image
reconstruction techniques (IRTs) such as  eclipse mapping (Horne 1985)
or  Doppler  tomography  (Marsh \& Horne   1988)  are able  to provide
spatially resolved  images of  accretion  discs on  a  micro-arcsecond
scale.
Since IRTs use the luminosity variation of the  disc with binary phase
during a considerable fraction of the  binary orbit, the reconstructed
images are time averaged pictures of the disc pattern.
This implies  that IRTs can reconstruct those  disc features which are
stable   over a period comparable to   a dynamical time   scale in the
co-rotating frame of the binary.

Numerical calculations by  various  authors (e.g.  Sawada,  Matsuda \&
Hachisu 1986, R\'o\.zyczka   \& Spruit 1993,  Heemskerk  1994, Larwood
1997,  Yukawa, Boffin \& Matsuda  1997, Stehle 1999) show that tidally
induced waves can result in two prominent spiral shock arms, which are
stationary on a dynamical time scale in a co-rotating frame.
The shocks are excited in the outer disc by the tidal action of 
the secondary star.
In  the binary potential field, the  disc gas revolves around the white
dwarf on non--axisymmetric  orbits where the streamline--density for a
pressure free gas  is highest towards the  secondary and lowest normal
to it (Paczy\'nski 1977).
In a low--temperature disc, the shape of the streamlines are still 
similar to the pressure free gas, but the variation of the
streamline--density with azimuth along a particle orbit will 
excite pressure waves which steepen to shock waves while they travel 
inward (Spruit, Matsuda, Inoue \& Sawada 1987). 
Additionally, shocks can be excited by the crossing of particle orbits
if the disc is larger than the tidal radius. 
The opening angle  of the shock arms depend   mainly on the  azimuthal
Mach number $M=v_{\phi}/c_{\mathrm{s}}$  with $v_{\phi}$ the azimuthal
disc gas  velocity and $c_{\mathrm{s}}$  the sound  speed, i.e. on the
rate with which the  shock is sheared by  differential rotation of the
disc gas  while the shock   travels inward radially  (R\'o\.zyczka  \&
Spruit 1989).

A strong radial viscous force,  if present in  accretion discs, may on
the other hand damp the shocks over a  short range or may even prevent
the pressure  disturbance to steep  into a  shock firsthand (Savonije,
Papaloizou \& Lin 1994).
As the origin of the viscosity is not known, we 
neglect in our calculations any radial viscous forces
and thus derive a maximum amplitude for the shocks.
If this is a sensible assumption is one of the main questions 
to be answered by comparing 
accretion disc models with yet to come detailed observations.

Despite  the firm prediction of shocks in accretion discs,
two  armed spiral  signatures have only
recently  been   found by  Steeghs,   Harlaftis \&   Horne  (1997) and
Harlaftis et al. (1999) in the eclipsing CV IP Pegasi during 
outburst. Doppler tomograms  revealed a strong two-armed spiral
pattern in the outer disc.
Even though Doppler tomography is ideally suited
to detect asymmetric disc structures (see e.g. Robinson, Marsh \& Smak
1993),  it is unclear  if previous observations failed  in
reconstructing spiral   arms because of  insufficient data  quality or
because spiral arms where  not present in the disc  at the time of the
observation.
To clarify   this question the  purpose  of this  paper is  to predict
observational signatures of  tidal density waves  in high  $M$ and low
$M$ accretion discs, the latter appropriate for accretion discs during
dwarf    nova outbursts. 

We use the  emission  line  model of  Horne  (1995) to  calculate  
emission line profiles for various
binary configurations, using  the   grid  of full  hydrodynamic   disc
calculations by Stehle (1999) as underlying models.

The  paper is organized  as follows:  in  Section \ref{models} we will
introduce  our time  averaged model  calculations  and our method of
constructing Doppler tomograms and line profiles.  In Section
\ref{ideal_tomo} we  present model predictions for several relevant
cases.
The effect of instrumental resolution and signal to noise on the 
detectability of tidal structure is explored in Section \ref{tomo_obs}.
We conclude the  paper in Section  \ref{discussion} with a  discussion
and summarize our results in Section
\ref{outlook}.

\section{Emission lines from spiral shocks}
\label{models}

\subsection{The grid of hydrodynamic disc models}

As  our  underlying   disc  models we  use  the   grid  of
geometrically thin 2D--disc calculations  by  Stehle (1999).
The  disc  thickness  is  explicitly   followed  by two   additional
equations in a one--zone model (see  Stehle  \& Spruit  1999 for a  detailed
description).
The spatial and   temporal evolution of  the disc  temperature follows
tidal   and  viscous heating,   the latter  in the  $\alpha$--ansatz of
Shakura \& Sunyaev (1973), as well as radiation from the disc surfaces.

\begin{figure}
\psfig{figure=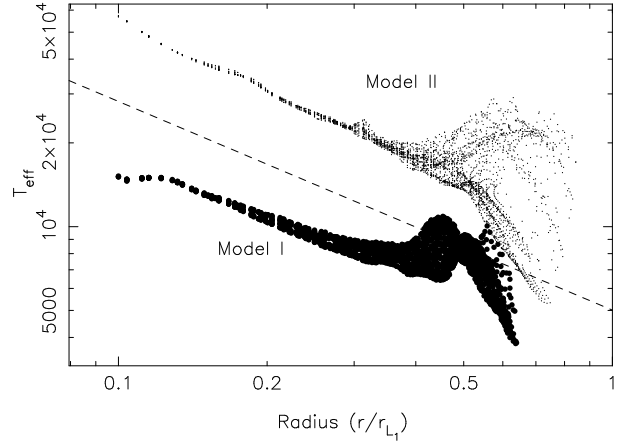,angle=-90,width=8cm}
\caption{The 
effective temperature $T_{\mathrm{eff}}(r)$ versus the disc radius
$r/r_{\mathrm{L_1}}$  for models I ($\alpha=0.01$,  lower line) and II
($\alpha$=0.3, top line).   $T_{\mathrm{eff}}(r)$ follows  closely the
power   $r^{-3/4}$ (dashed line),  as    expected for hot,  stationary
accretion discs.  Near   the outer  disc  edge, the   effects of tidal
heating   and non-axisymmetry   result   in  a   wide  range of   disc
temperatures.}
\label{trplot}
\end{figure}

Our time dependent calculations were time averaged over at least 
one binary orbit in the co-rotating frame of the binary, i.e.
\begin{equation}
\bar{f} = \frac{1}{t_{\mathrm{f}} - t_{\mathrm{i}}}
\int_{t_{\mathrm{i}}}^{t_{\mathrm{f}}} \, f(t) \, 
{\mathrm{d}}t
\end{equation}
where $f$ is any conservative physical quantity, like
the momentum in radial direction $p_{\mathrm{r}}$,
the angular momentum $p_{\phi}$, 
the vertical momentum $p_{\mathrm{z}}$,
the surface density $\Sigma$ or
the internal energy density $e$.
Other quantities, as e.g. the disc gas velocity, are derived from the
time averaged values.
We  will  deal further on only  with  time averaged values  and thus we
subsequently drop the bar in our notation for convenience.
By averaging our  models in time, only  those disc features which  are
reasonably stable over a dynamical time scale will still be visible.

We will be  presenting the  results of   two model simulations.   Both
models are  simulations  of the accretion  disc in  a binary with mass
ratio $q=M_2/M_1=0.3$   and   orbital  period   $P=2.3$   hours.   The
$\alpha$--type Shakura \& Sunyaev viscosity in model  I is 0.01 and we
time-averaged over $\sim$1.5 binary orbits.
Model  II represents a typical
dwarf nova outburst disc with  $\alpha=0.3$,  time averaged over $\sim$
one
orbital period.
We also calculated model predictions for various other mass ratios and
different averaging  times   which  all  share the   same  qualitative
features since the spiral arms are very stable in the binary frame and
mainly depend on the disc  temperature (see also  Stehle 1999).  These
two models in particular were selected as a typical comparison between
a  small,  cool disc   (I)   versus a  large, hot  viscous   disc (II)
illustrative for the discs  in CVs.  We show  in Fig. \ref{trplot} the
run  of  $T_{\mathrm{eff}}$  with  disc  radius  $r$ in   units of the
distance      to    the  $L_{1}$         point.    In   both     cases
$T_{\mathrm{eff}}(r)\propto r^{-3/4}$ in the  inner parts of the disc,
as  expected  from stationary viscous  accretion  discs.  Close to the
outer  disc edge though, tidal  heating  and non-axisymmetry result in
the large spread of $T_{\mathrm{eff}}(r)$.

\subsection{Doppler tomograms and line profiles}

For our analysis we use a right--handed Cartesian coordinate system 
$(\vec{e}_{\mathrm{x}}, \vec{e}_{\mathrm{y}},\vec{e}_{\mathrm{z}})$  
centred on the white dwarf, with the positive X--axis in the direction of
the secondary star and $\vec{e}_{\mathrm{z}}$ parallel to the rotation
vector of the accretion disc.
The disc gas velocities are transferred
to inertial velocities corresponding to the centre of mass
of the binary.
To  calculate the line emissivity  for each disc  grid cell, we follow
the lines of Horne \& Marsh (1986) and Horne (1995). The emission line
surface brightness $J$ is obtained  by integrating over the local line
profile times the foreshortening due to the orbital inclination $i$;
\begin{equation}
J=\cos i \int S_L(1-\exp{(-\tau_{\nu})})~ d\nu
\end{equation}
In the  case of optical  thin line emission,  $J(x,y)$ is proportional to   
the surface density $\Sigma(x,y)$ 
and the local line source function $S_L(\lambda)$.
However, most strong emission lines such as the hydrogen Balmer series
are optically thick and saturated. The  area under the profile is then
well approximated by  the product of $S_L$  and the frequency interval
over which the optical depth exceeds unity.
Using eq. (15) from   Horne (1995), the  local  line emissivity  of  a
saturated  emission line with  rest wavelength $\lambda_0$  is in that
case given by:
\begin{equation}
J(x,y) = S_L \frac{\Delta V}{\lambda_0} \cos i \sqrt{8 \ln{\tau_0}}
\end{equation}
$\Delta V$ is the total  velocity dispersion along  the line of sight,
$\tau_0$ the optical  depth of the centre  of the emission line. If we
furthermore assume  local thermodynamic equilibrium, $S_L$  equals the
Planck function.
The velocity dispersion consists of a thermal component 
$V_{\mathrm{th}}=\sqrt{kT/m_{\mathrm{H}}A}$, 
with $m_{\mathrm{H}}$ the mass of a hydrogen atom and
$A$ the atomic weight,
and a shear term $V_{\mathrm{shear}}$,
derived from the full, non-Keplerian, velocity field.
$\Delta V$ is evaluated for each grid cell and viewing angle
according to
\begin{equation}
\Delta V^2 = V_{\mathrm{th}}^2 + V_{\mathrm{shear}}^2
\end{equation}
where
\begin{eqnarray} 
V_{\mathrm{shear}} &  = & \frac{\Delta Z}{\cos{i}} 
\: \vec{e} \cdot \vec{\nabla}\vec{V} \cdot \vec{e}\\
                   & = &  \frac{\Delta Z}{\cos{i}} 
\:\vec{e} \cdot 
\left( \begin{array}{ccc}
\frac{\partial V_x}{\partial x}&
\frac{\partial V_y}{\partial x}&
\frac{\partial V_z}{\partial x}\\
\frac{\partial V_x}{\partial y}&
\frac{\partial V_y}{\partial y}&  
\frac{\partial V_z}{\partial y}\\ 
\frac{\partial V_x}{\partial z}&\
\frac{\partial V_y}{\partial z}& 
\frac{\partial V_z}{\partial z} 
\end{array} \right) \cdot \vec{e}
\end{eqnarray}
$\vec{e}$ denotes the earth vector pointing from the grid cell towards
the observer, and   $\vec{\nabla}\vec{V}$ is the  tensor of  the local
velocity gradient including  all  anisotropic terms.  In  this way the
velocity field of our simulations self consistently provides the total
velocity dispersion along  each line of  sight. $\Delta Z$ denotes the
vertical extent  of the emission line  layer and is assumed  to be $ 0
\leq \Delta Z \leq H$.

\noindent Since $\tau_0 \propto \Sigma/\Delta V$ we derive:
\begin{equation} 
J(x,y,\phi,i) \propto B_\nu(T_{\mathrm{eff}})~ \Delta V(x,y,\phi,i)
\sqrt{\ln(\frac{\Sigma(x.y)}{\Delta V(x,y,\phi,i)})}
\end{equation}
The line  emissivity   is  mainly  sensitive  to   the local  velocity
dispersion $\Delta V$.  The  large gradients across the spiral  shocks
will therefor enhance the  local emissivity of saturated
lines significantly at the location of the spiral arms.

The line emissivity is transformed from position $(x,y)$ to velocity 
coordinates $(V_x,V_y)$ by multiplying with
the Jacobian of the coordinate 
transformation, i.e.
\begin{equation}
 J(V_{\mathrm{x}},V_{\mathrm{y}}) = 
\left| \begin{array}{cc} \frac{\partial V_x}{\partial x} &
\frac{\partial V_x}{\partial y}\\ \frac{\partial V_y}{\partial x} &
\frac{\partial V_y}{\partial y} \end{array} \right|^{-1} 
J(x,y).
\end{equation}

For  each binary  phase   $\phi$  the  line   profile $f(v,\phi)$   is
synthesised  by adding up   the  individual contributions of the  grid
cells   where  the  emission   line  contribution is   Doppler shifted
according   to the  local  gas   velocity  along  the line of   sight,
$V_{dop}=-\vec{V}  \cdot \vec{e}$. We  adopt the usual convention that
phase  0.0 corresponds to  inferior conjunction of the companion star,
i.e.  mid-eclipse in high inclination systems.
The visibility of each grid cell is tested for eclipses
by the companion star which we assume to fill its critical Roche volume.
Self-shadowing of the disc, important for high--inclination
CVs due to the vertically extended disc rim will be discussed 
elsewhere (Stehle \& Steeghs, MNRAS, in preparation).

The  above description  to    construct emission line  profiles  holds
principally  for all  {\it  saturated} emission lines.  We  choose the
Hydrogen   $\beta$ line  at 4861\AA~  as   our reference line for most
calculations. The  Balmer lines  are the  most  prominent lines in  CV
spectra and allow disc  reconstructions at a high  signal to noise. It
serves as a typical representative line  observed in optical
emission line  studies of   CVs.  Since we    are concerned  with  the
dynamical  properties of the  line  profiles  rather than  calculating
absolute  line strengths, this particular choice   does not affect our
subsequent analysis. 
To   illustrate the effects of saturation   we will calculate both the
case of no shear and maximum shear broadening as  well as compare with
the properties of a heavy element such as Calcium.

\section{Model predictions}
\label{ideal_tomo}

In  this section   we present  predicted emission   line  profiles and
Doppler tomograms for the disc models  limited only by the finite grid
resolution of  our hydrodynamical   calculations.  As we will see   in
Sec. \ref{tomo_obs},  the limits  set   by the   instrumentation  will
dominate over  our limited numerical  grid  resolution and it  is thus
appropriate  to  call our tomograms   in   this Section ``ideal''.  We
identify the signatures   of spiral waves   in the  emission lines  of
accretion discs.
As the  global shape of  the spiral shocks  is characterised by the Mach
number of the disc,  we  present our model predictions   in
descending order of $M$. 

\begin{figure*}
\centerline{\psfig{figure=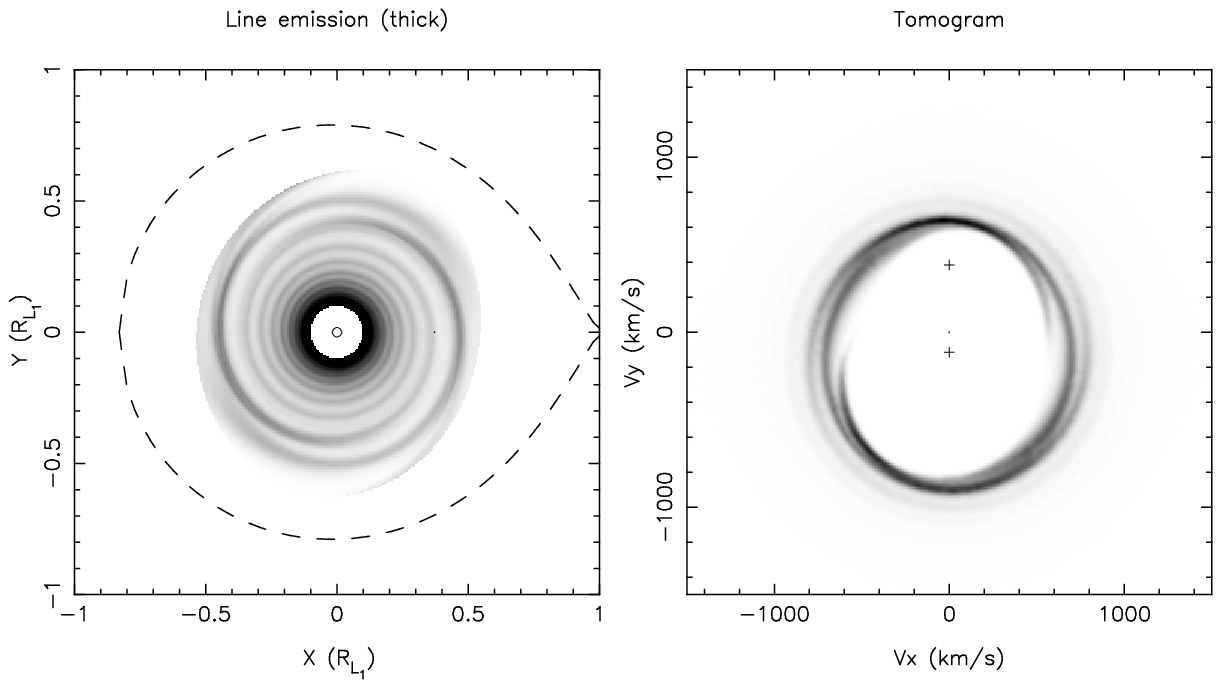,width=17cm}}
\begin{tabular}{cc}
\psfig{figure=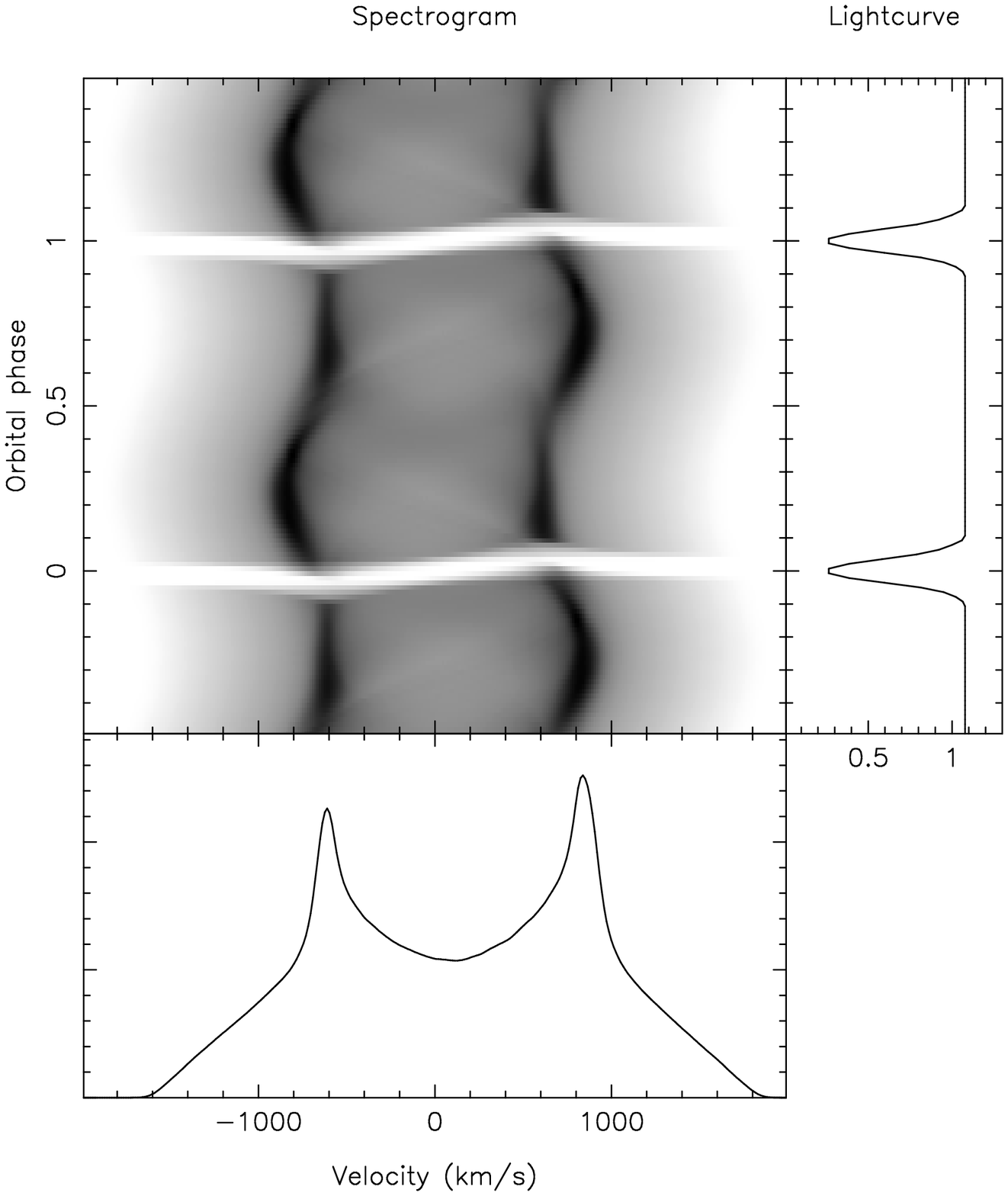,width=8.5cm}&
\psfig{figure=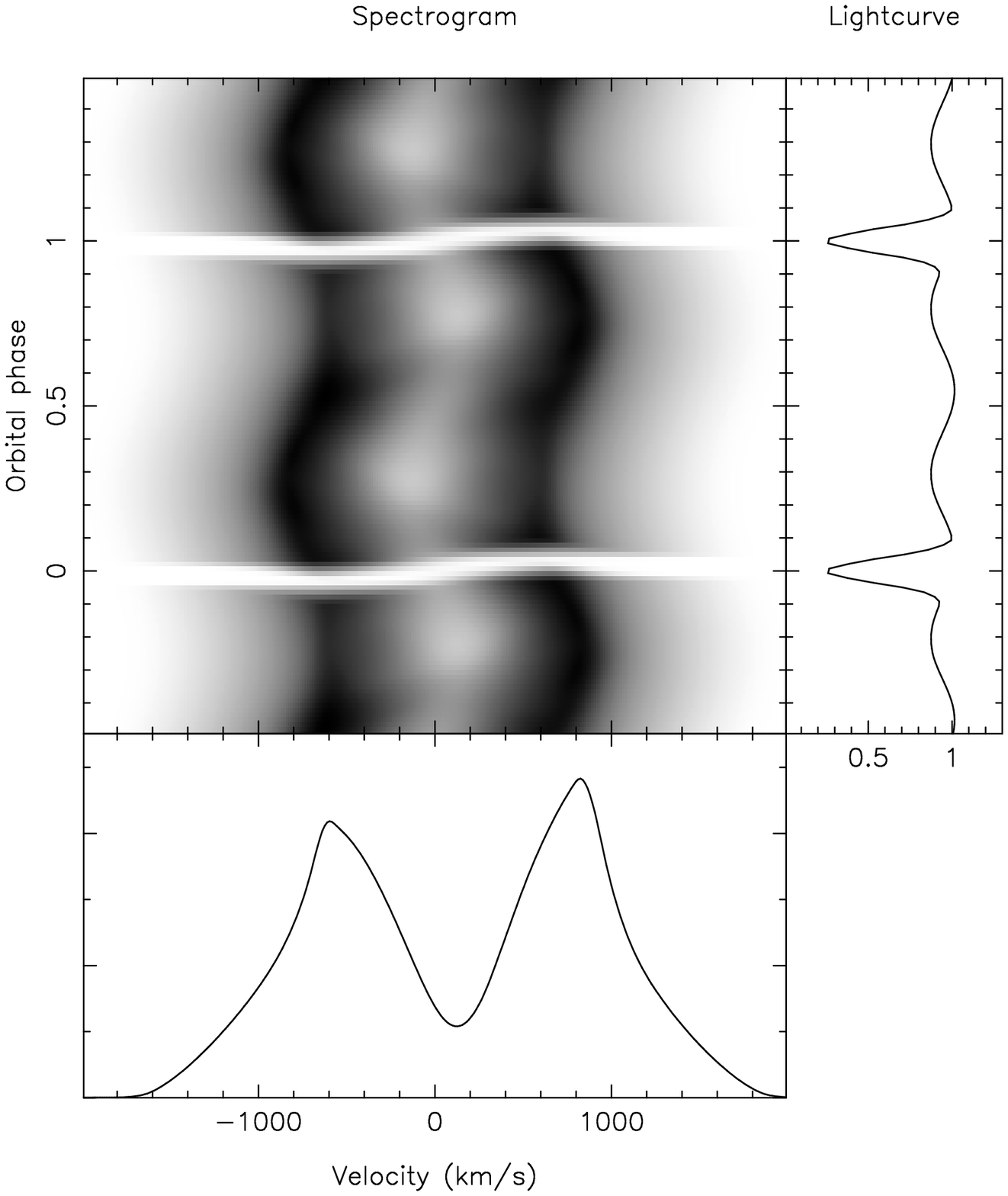,width=8.5cm}\\
\end{tabular}
\caption{Model I: 
an  $\alpha=0.01$ disc in a   binary with $q=M_2/M_1=0.3$.  Top panels
show  the distribution  of line  emission  in both spatial coordinates
(left) as well as  in the $V_xV_y$-plane. The  crosses in the tomogram
denote the projected radial velocity of the  white dwarf (lower cross)
and secondary star. Bottom panels display the calculated emission line
profiles for Hydrogen-like  lines at  an inclination of  80$^{\circ}$.
Bottom  left with  no  shear broadening  (i.e. a   thin  emission line
layer),  right with maximum   shear  broadening when line emission  is
produced  across the whole disc height.  To the  right of each trailed
spectrogram  is the  total line  flux  light curve,  and below it  the
emission line profile at orbital phase 0.75.}
\label{quiescence}
\end{figure*}

\begin{figure*}
\begin{tabular}{cc}
\psfig{figure=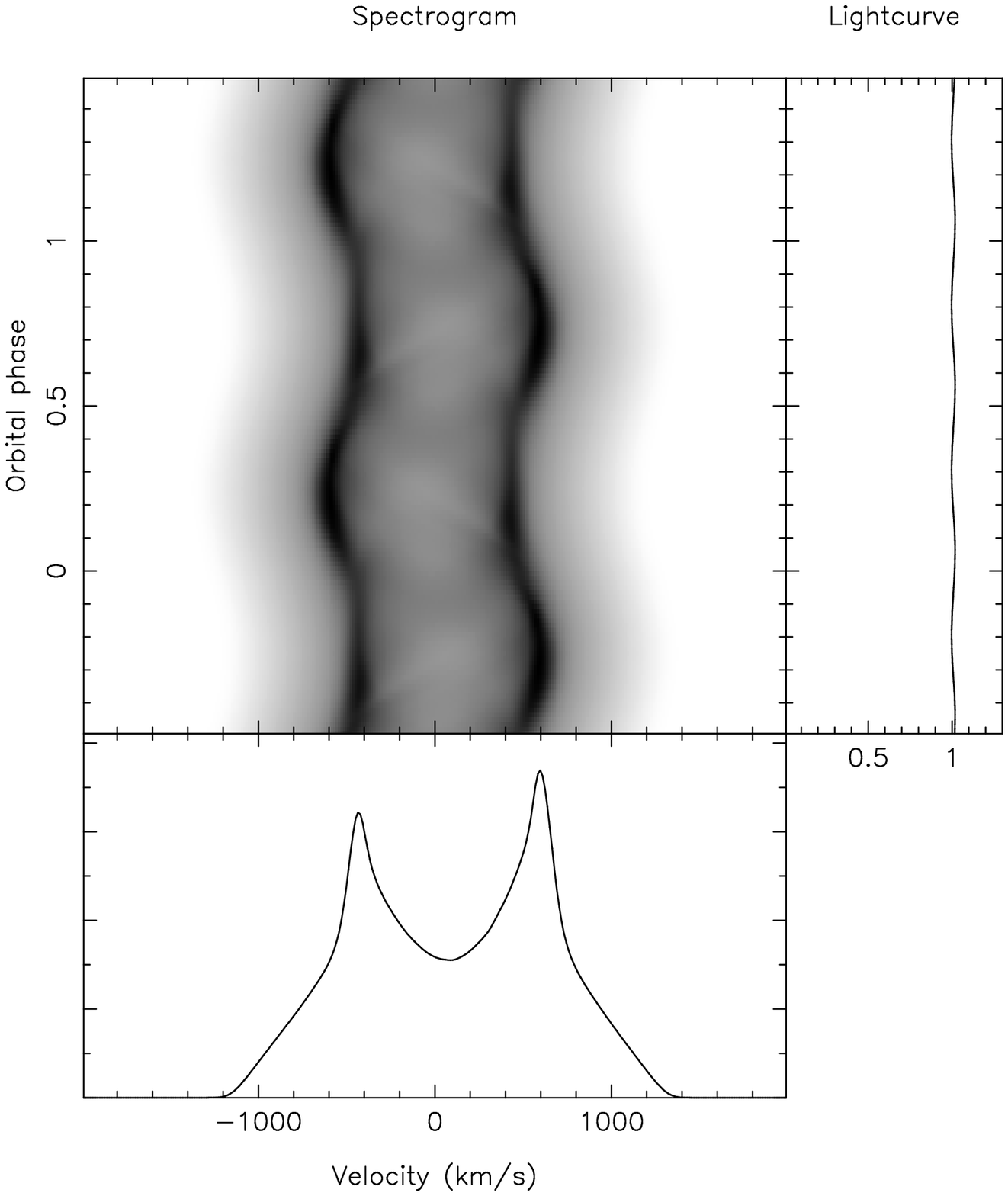,width=8.5cm}&
\psfig{figure=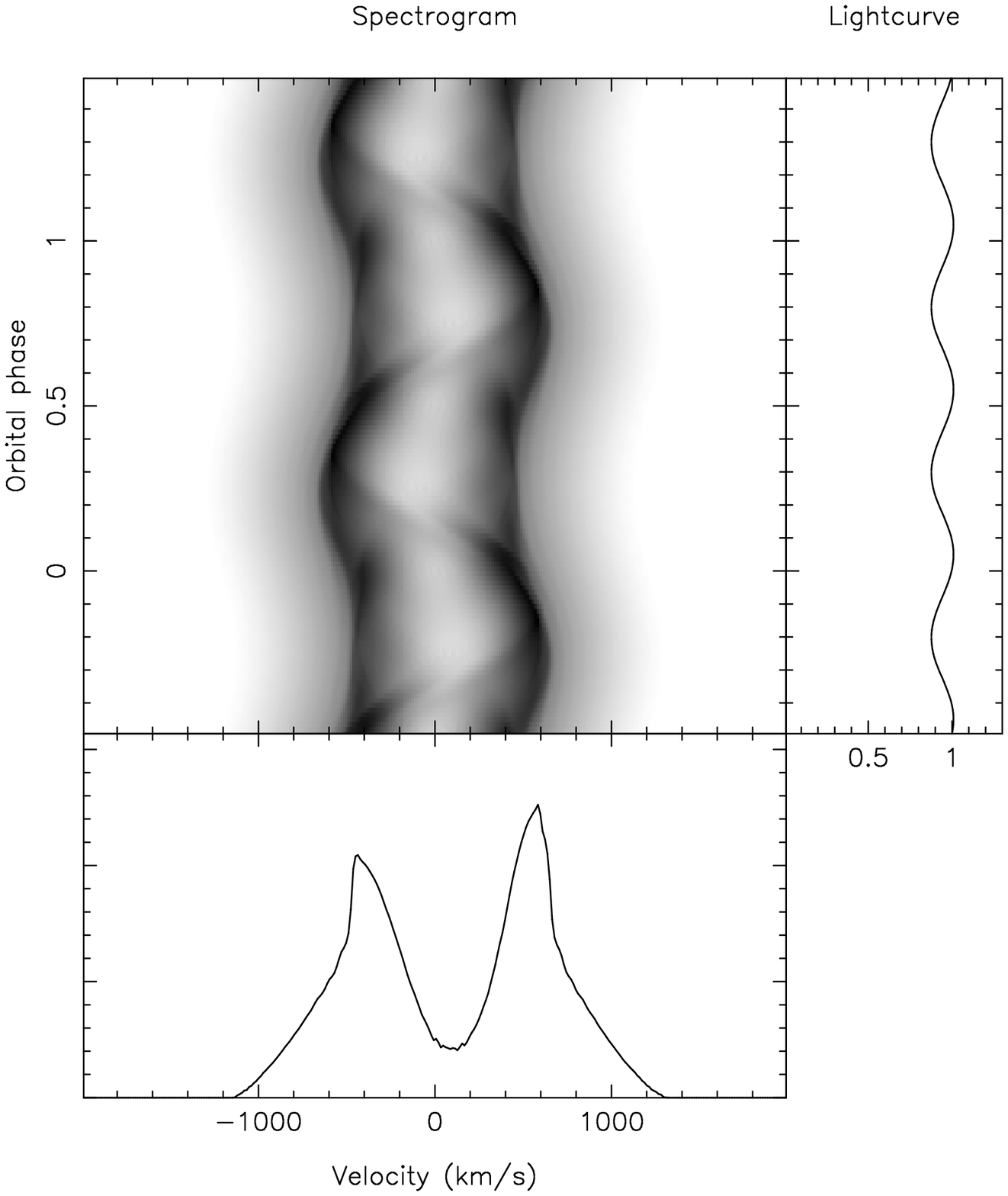,width=8.5cm}\\
\end{tabular}
\caption{The 
effect of atomic weight on  the shear broadened profiles. Left are
the   line  profiles   for Model  I   viewed   at an  inclination   of
$i=45^{\circ}$   for a  Hydrogen line  (A=1).   Right panel  is  for a
saturated Calcium  line (A=40). Because   of the much smaller  thermal
velocity in the case of Calcium, the shear broadening is much stronger
than for Hydrogen  lines. This  results  in  a stronger signature   of
spiral waves in the emission lines.}
\label{45deg}
\end{figure*}

\subsection{Axisymmetric discs}

The properties of   emission lines from axisymmetric  accretion  discs
have been  investigated by several  authors (e.g.  Smak 1981, Horne \&
Marsh 1986), of which we summarise a few relevant features.  The large
Kepler velocities of a few thousand km s$^{-1}$ in the inner disc down
to several hundreds km s$^{-1}$ in the  outer disc, produce very broad
emission line profiles,  decoupling the global  shape of the line from
the local line radiative transfer.  A  double peak shape results where
the peak velocities  reflect the motion of the  gas at  the outer disc
edge, which in turn can be used to estimate the  size of the disc if a
velocity field, such as Keplerian, is assumed.

The orbital velocity  of  the white dwarf  is sometimes   estimated by
measuring the centroid of the observed line profile, which should show
a sinusoidal  dependence with orbital phase  as the white dwarf orbits
around the  binary center of mass.  While  the line profile as a whole
thus moves,  the separation between the  double peaks is constant with
phase.   In  the velocity   coordinate  space  of Doppler   tomograms,
axisymmetric discs will produce axisymmetric circular images, centered
on the  radial velocity  of the   white dwarf ($V_x=0,V_y=-K_1$,  with
$K_1$  the projected orbital velocity  of  the white dwarf). For  each
radius, a corresponding ring  at the local  disc velocity is produced,
with the smallest velocities corresponding to the outer edge.

\subsection{Model I: High Mach number disc}

In Fig. \ref{quiescence} we show the model predictions 
for the high Mach number CV disc of model I, 
viewed at an inclination of 80$^{\circ}$. 
The  spatial distribution of the line  emission  is plotted in the top
left panel and displays a rather small disc, varying in radius between
$r_{\mathrm{out}}(\phi) \simeq 0.55-0.65\, r_{\mathrm{L_1}}$, with two
tightly wound spiral shock arms covering the disc down to small radii.
If we plot the same distribution in  velocity coordinates (top right),
the typical ring  shape appears, with   the two shock arms  visible as
sharp arcs  of   enhanced line  emissivity   in  the  top  and  bottom
areas. The outer  disc edge provides a  cut off  of any emission  with
velocities lower than $\sim$ 600 km s$^{-1}$ while at high velocities,
the inner disc, the tomogram becomes more symmetrical.
The emission line profiles  in the corresponding  trailed spectrograms
(Fig.    \ref{quiescence},  bottom  panels) feature   two sharp peaks,
reflecting the velocities of the outer disc gas. Two cases are plotted
to illustrate  the effect   of shear  broadening  due  to the   finite
thickness  of the emission line region  (Horne \& Marsh 1986).  In the
bottom left panels, the emission line layer is assumed to be very thin
($\Delta Z/H \simeq  0$), so that  shear broadening effects vanish and
sharp and narrow double peaks are formed. This  is very similar to the
optical  thin case.   On the  right   is  the case   of maximum  shear
broadening, assuming the  emission  lines are  formed across  the full
disc thickness such that $\Delta Z = H$  in eq. (5).  Broad peaks with
a V-shaped, rather than U-shaped valley between the peaks are formed.

The  presence of  spiral arms  slightly distorts the  velocity  of the
peaks     as   a   function    of    binary  phase,   varying  between
$v_{\mathrm{peak}}(\phi)=$ 690 -- 750 km  s$^{-1}$ with respect to the
white dwarf.   On average,   the  double peak separation  is  $\langle
v_{\mathrm{peak}}\rangle=725$ km s$^{-1}$.  This corresponds to a disc
radius    of     $r_{\mathrm{out}}         \simeq         GM_1/\langle
v_{\mathrm{peak}}\rangle^2   =   0.61  \, r_{\mathrm{L_1}}$,  assuming
Keplerian velocities.  While the  disc  is slightly tidally  distorted
and  elongated along  the $Y$-direction,  the assumption of  Keplerian
velocities still provides a reliable estimate of the average disc size
in this case.

Shear  broadening can    be     particularly important  for      heavy
elements. Their   low  thermal velocity results   in  a stronger local
enhancement of the line flux due  to velocity gradients. To illustrate
this, Figure  \ref{45deg} shows the  line profiles of a  light element
(A=1)  versus     that  of  Calcium (A=40)   at     an inclination  of
45$^{\circ}$. While for hydrogen like lines the shear broadening is no
longer  important due  to foreshortening,  the   Calcium line displays
strong anisotropic components originating along the spiral arms.  This
increases the double peak  variation from 8\% in the  case of no shear
to 13\% for maximal shear in heavy lines such as Calcium.

The  orbital variation in  the separation of  the double peaks, is the
predominant  signature  that  the  disc  is asymmetric  and  that  the
emissivity  is  not  uniform.  In   the   case of  considerable  shear
broadening, the line emission from  the spiral shocks is  additionally
non-isotropic resulting in a  binary phase dependent variation  of the
peak strengths with a maximum at phases 0.25 and 0.75 as is visible in
the emission line flux light curves to the  right of the spectrograms.
  Such non-isotropic
line emission cannot be modelled by Doppler tomography, since only the
average  line flux  at  a  given velocity   is  provided, but  can  be
identified in the emission  line profiles directly  as a  signature of
shocks in  the outer disc.  While  saturated emission lines from heavy
elements are thus  good indicators of  large velocity gradients,  they
are  unfortunately usually   weak lines  and  do  not   provide
sufficient signal to noise for Doppler imaging in existing data sets.

\subsection{Model II: dwarf nova outbursts}

\begin{figure*}

\centerline{\psfig{figure=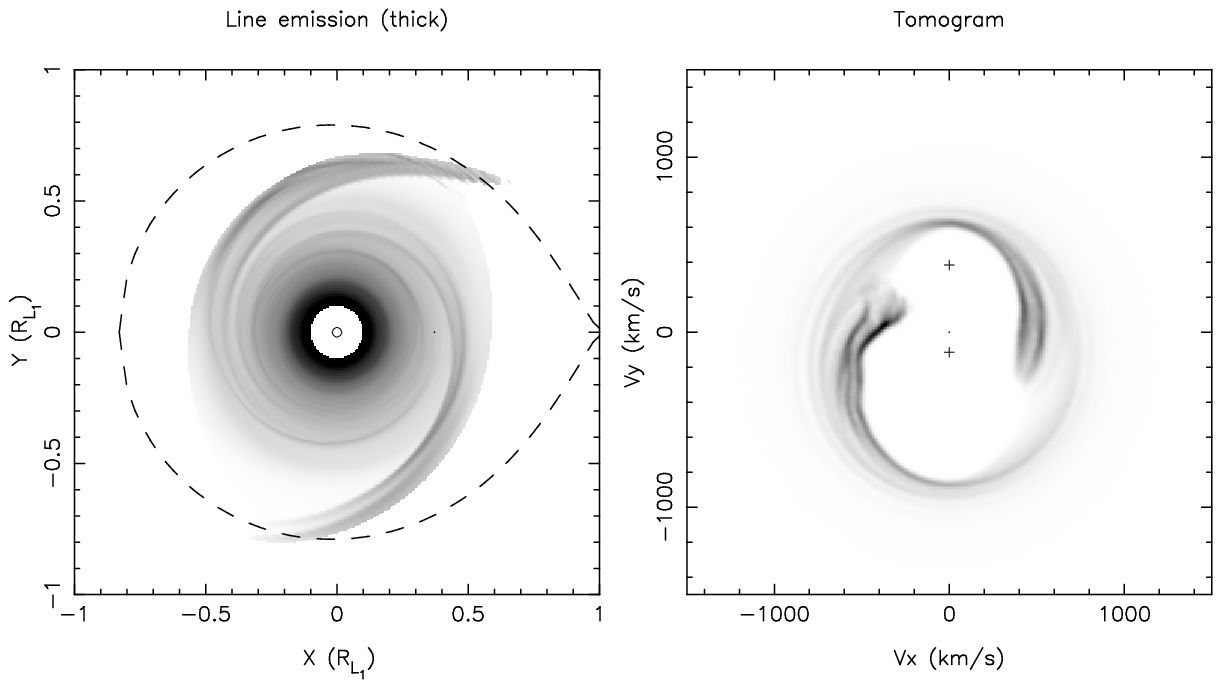,width=17cm}}
\begin{tabular}{cc}
\psfig{figure=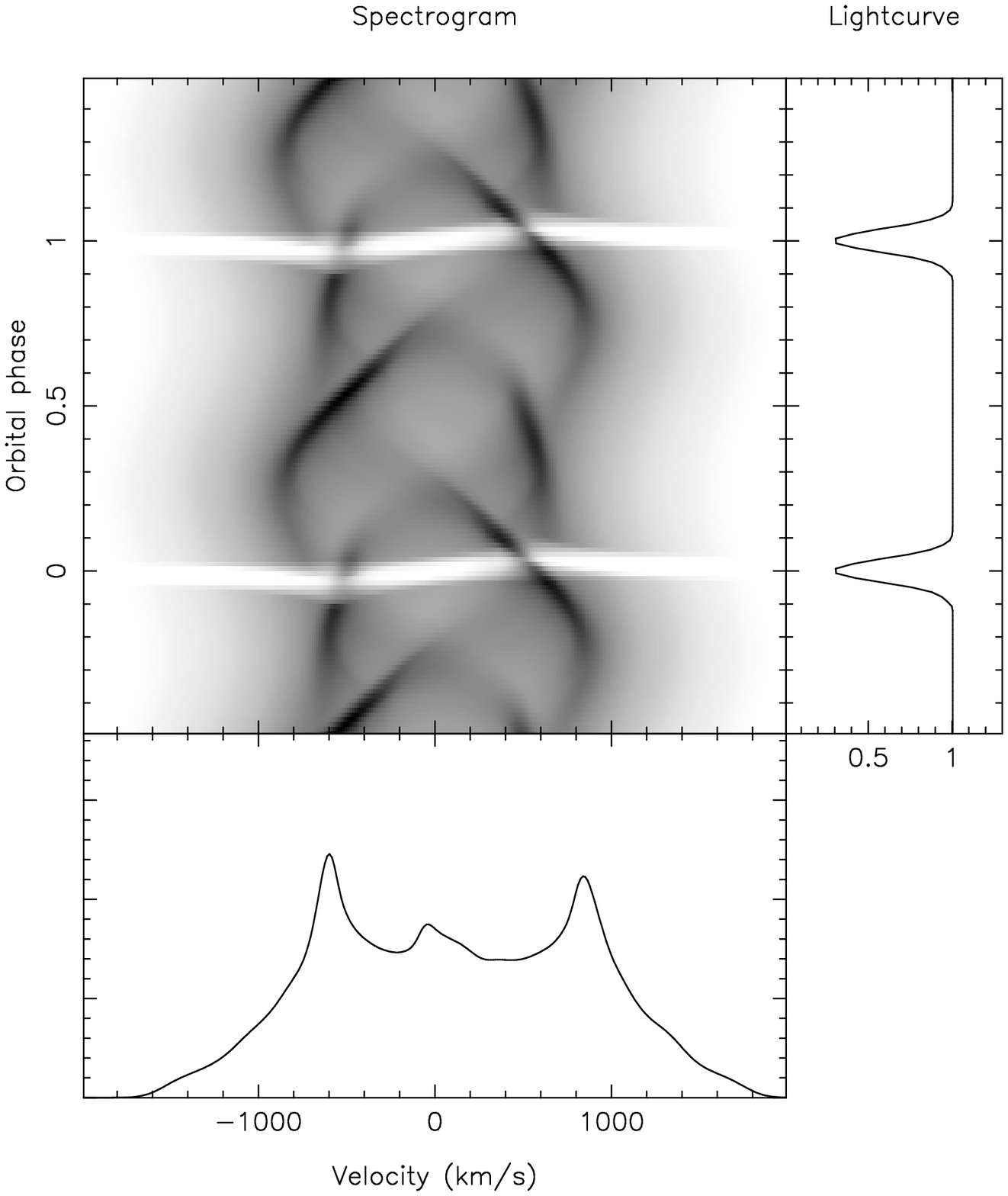,width=8.5cm}&
\psfig{figure=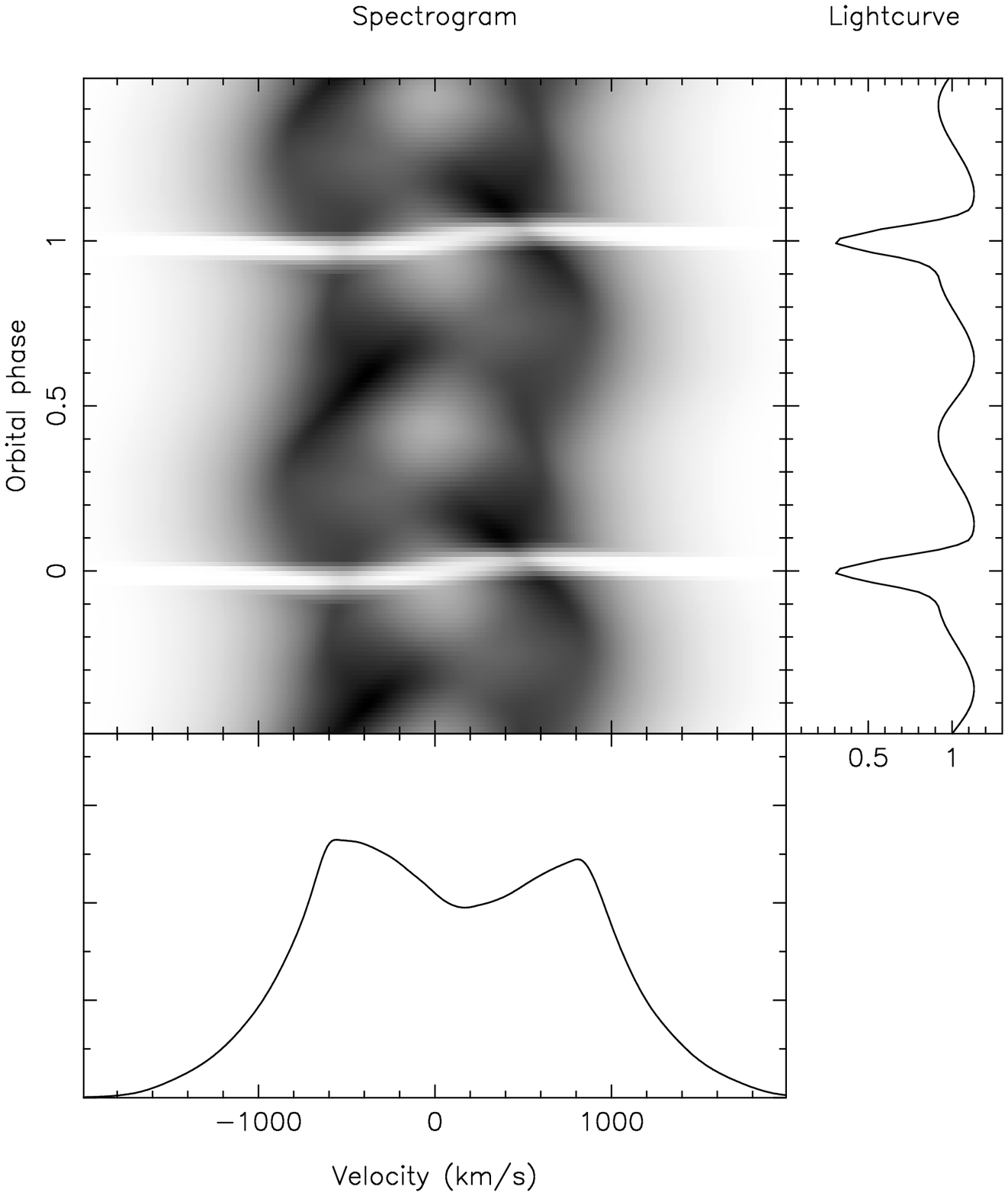,width=8.5cm}\\
\end{tabular}
\caption{As Fig. {\protect{\ref{quiescence}}} but now for the low
         Mach number disc of model II.}
\label{outburst}
\end{figure*}

By  increasing   the  parameter   of  the small    scale viscosity  to
$\alpha=0.3$ in Model II, viscous  spreading becomes more efficient in
pushing     the    accretion  disc    towards       larger radii    of
$r_{\mathrm{out}}(\phi) \simeq 0.6-0.8 \, r_{\mathrm{L}_1}$ (see Fig.
\ref{outburst}, top left
panel).   The increased local viscous  heating additionally  produces
higher
disc temperatures, and  thus lower Mach  numbers, now of the  order $M\sim
5-20$. This  is reflected in the open geometry
of the two spiral shock arms in the outer disc.
In Fig. \ref{outburst} we present the calculated emission line
properties for model II, again at an inclination of $i=80\degr$.

\begin{figure}

\psfig{figure=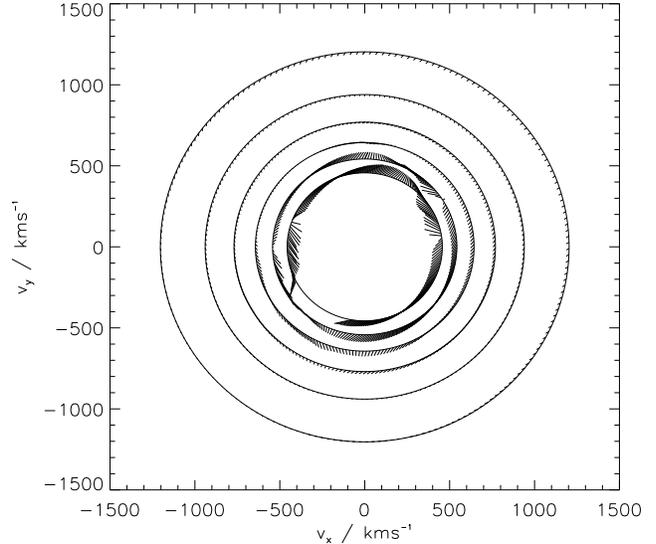,width=8cm}
\caption{Non-Keplerian  
velocities in model II. The   circles denote the Kepler
velocity at  a few radii  in the  disc. The small  ticks point  to the
actual    disc velocity   at that   radius   and  azimuth.  The  tidal
distortions become prominent in the outer disc, i.e. at low velocities
in the inner regions of the Doppler maps.}
\label{distortion}
\end{figure}

It is difficult to assign a meaningful disc radius as
the line emissivity is predominantly localized at the two spiral arms.
The strong $m$=2 spiral wave  pattern in the  outer disc dominates the
corresponding    Doppler tomograms, shown   in  the  top right panel of
Fig.  \ref{outburst}.  The   line  emission  extends    towards  lower
velocities ($\sim$  500  km s$^{-1}$), partly because  of  the increased
disc
size,  and partly because sub-Keplerian    motions are significant  in
these outer regions. Figure
\ref{distortion} shows the deviation from Keplerian velocities at a few
representative  radii. While the disc   is  close to Keplerian in  its
inner regions (high velocities), 
the departures at the outer disc near $ v \sim $ 500 km s$^{-1}$ are
significant and of the order of $\sim$100 km s$^{-1}$.

The line profiles  are   easily distinguishable from   an axisymmetric
disc.  The $m=2$  spiral pattern results  in converging  emission line
peaks, with a cross over near phases 0.25 and 0.75.  This contrasts to
the axisymmetric case,   with   a constant  double   peak  separation,
independent of  orbital phase.  Though a  non-circular disc could also
produce a variation in the double peak separation with phase, it would
not produce  the existence of multiple  peaks as one crosses  from one
spiral  arm to the next, with  a corresponding sudden jump in velocity
of several  hundred km s$^{-1}$, a unique  signature of a spiral shock
pattern.  Again the  large  velocity gradients are important   at high
inclinations where shear   broadening can dominate.  This  produces  a
variation in the emission line light curve (Fig. \ref{outburst} bottom
right panel) which is considerably larger compared to Model I.

\section{Reconstructing spiral structures in CV discs}

\label{tomo_obs}

\subsection{Realistic data quality}

In the previous section we discussed our  model predictions limited by
the grid resolution  of  our hydrodynamical  models.  We  now  include
instrumental  noise   and  finite spectral/time   resolution  in   our
predicted  data to provide  a realistic comparison  with observed data
and constrain the necessary  data  requirements to  reconstruct  tidal
arms in CV discs.

Three instrumental effects     play  a role.   First  the   wavelength
resolution of  the individual    spectra,  setting  a limit  to    the
velocities that  can be resolved.  The   emission lines typically span
several thousands of km  s$^{-1}$, which need to  be sampled  by ample
number of pixels. On the other hand, resolutions  much better than the
width of the  local  line  profile ($\sim$  10  km s$^{-1}$)  are  not
providing any additional information.  A resolution of $\sim$ 20-80 km
s$^{-1}$ is therefor usually optimal, though is practically limited by
signal to noise requirements.

Secondly, the  phase or time  resolution has   to be short  enough  to
provide sufficient phase sampling   across the binary orbit   to avoid
artifacts in the image reconstruction (see also  Marsh \& Horne 1988).
The higher the resolution of the  input data, the more projections are
required to   provide  sufficient sampling  and  exploit the available
resolution.  To avoid  under sampling the image  one typically needs at
least 50 orbital phases. 

Finally, the signal to noise of  the individual spectra will determine
how well the data  will constrain the  image structure. In the case of
maximum entropy reconstruction as is the  case here, the reconstructed
image is a balance  between fitting the  data using maximum likelihood
statistics and selecting the simplest  image using the maximum entropy
criterion.

\begin{figure*}

\psfig{figure=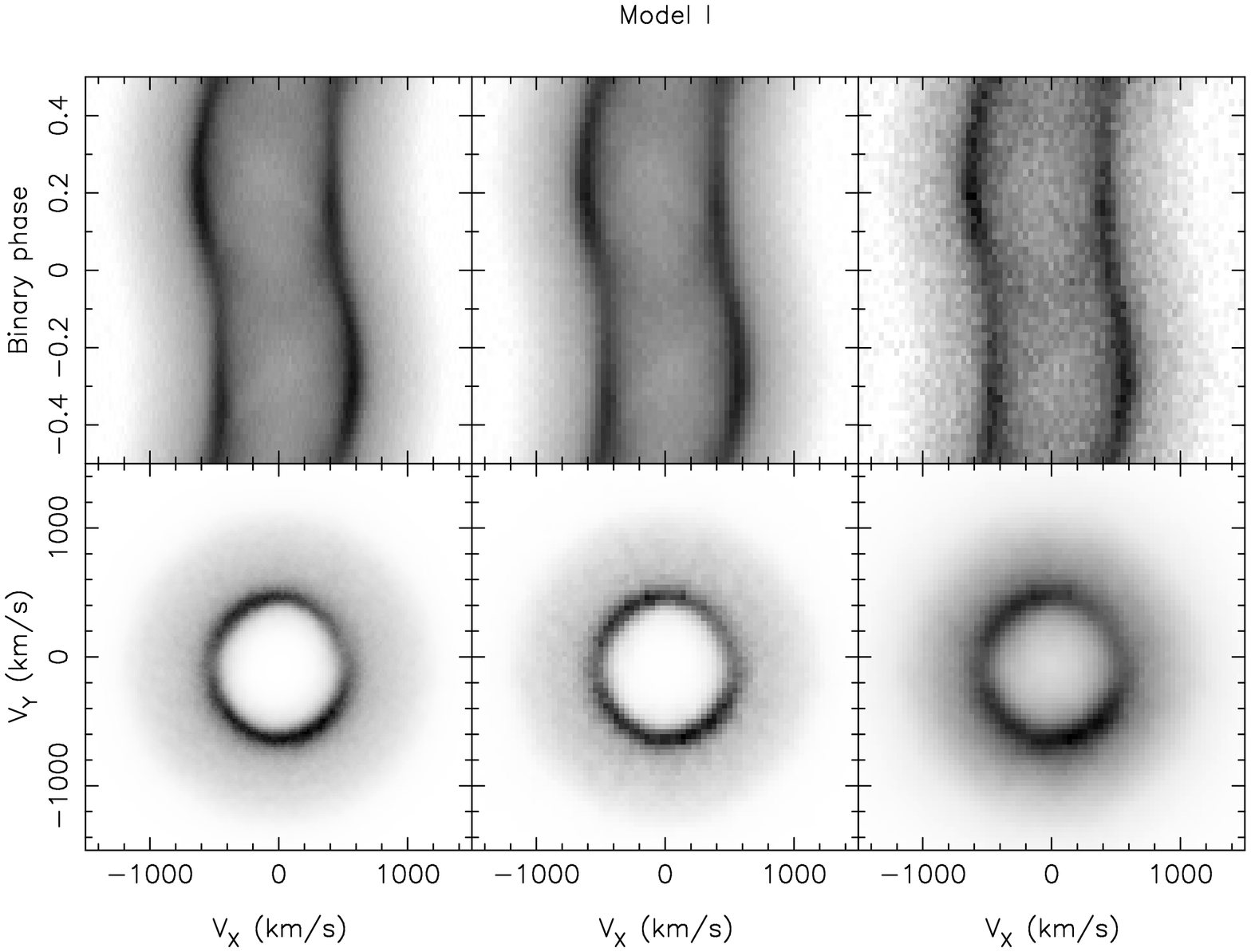,angle=-90,width=17cm}

\caption{Tomograms and trailing spectra of the high 
Mach number simulation (model I) for a strong emission line. 
From left to right 
         the  2-pixel wavelength/velocity resolution
         is (40, 80, 80) km s$^{-1}$ and the ratio of signal 
         to noise is (50, 50, 15) in the continuum.}
\label{5recon}
\end{figure*}

\begin{figure*}

\psfig{figure=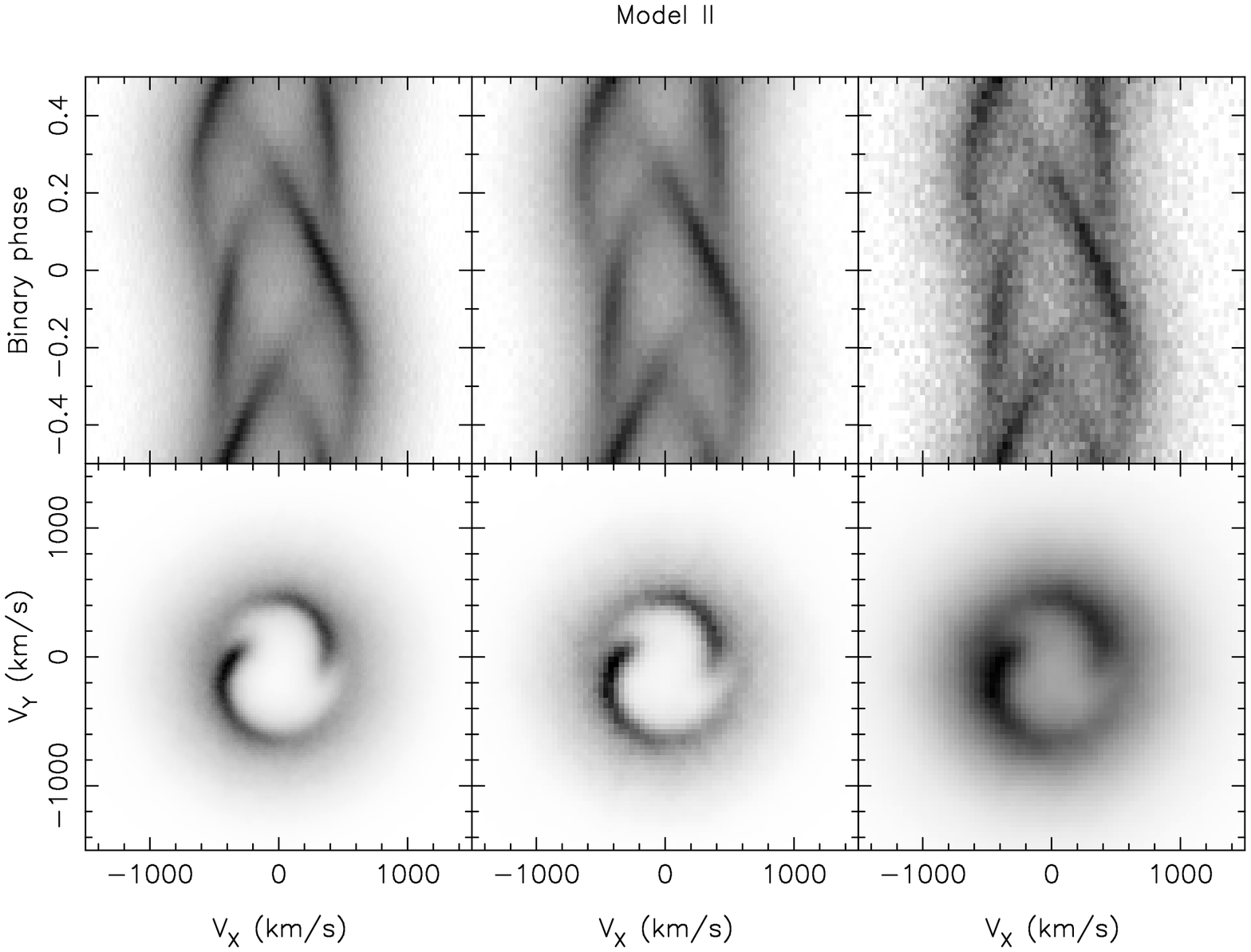,angle=-90,width=17cm}

\caption{As Fig. \protect{\ref{5recon}} but for model II}
\label{7recon}
\end{figure*}

To test the   importance  of above   mentioned effects, we   start  by
calculating  ideal  line profiles   (such   as  presented  in  Section
\ref{ideal_tomo}) for an inclination of 45$\degr$ at 50 orbital phases
equally spaced over the binary orbit.  The data is then convolved with
a Gaussian with FWHM equal to the  desired instrumental resolution and
binned in  wavelength such that 2 data  pixels  cover the instrumental
profile.   Poisson noise    was  subsequently  introduced  to  achieve
different signal to noise levels in the  line profiles as follows.  We
added a  continuum to our line profiles   such that the  line strength
(line - continuum)   equals about  two times  the  continuum  level, a
typical value  for the strong emission  lines in dwarf novae.  Poisson
noise was then added  to achieve the desired signal  to noise level in
the continuum. Using the  synthetic  noisy data, we construct  maximum
entropy  Doppler tomograms using  the same code  that  is used for the
observations
\footnote{DOPPLER package developed by Tom Marsh, 
see {\tt http://www.astro.soton.ac.uk/$\sim$trm/software.html}}

\subsection{High M model}

Figure \ref{5recon} shows the  image reconstructions based on model I,
at an inclination of 45$\degr$ and including maximum shear broadening.
The  top panels show  the  simulated  data,  the  left has  a  2-pixel
resolution of 40  km s$^{-1}$, the middle and  right have a resolution
of 80 km s$^{-1}$. The  signal to noise in  the left and middle panels
is $\sim$ 50 in the assumed continuum, while for the right panel it is
degraded  to $\sim$   15.   Below each spectrogram  the  reconstructed
Doppler tomograms are shown, all on the same scale.

It is difficult to reconstruct a  clear signature of the tightly wound
spiral waves.  Except  for an enhancement of  the disc emission in the
top left  and  lower  right, such discs  will  look  very  similar  to
axisymmetric discs,  even if high quality  data  is available. As data
quality  is degraded  the slight  variation  of the double  peaks with
binary phase becomes difficult to measure.  We conclude that we do not
expect to   see obvious evidence  of spiral  waves in  Doppler maps of
small, low $\alpha$ accretion discs even if tightly wound spiral waves
are present.

\subsection{Low M model} 

Figure  \ref{7recon} plots similar  reconstructions now based on model
II.   Resolution  and signal to   noise levels  are  identical to  the
previous model for comparison.  The strong two armed spiral pattern is
readily recovered  even at low signal  to  noise and medium resolution
(Fig. \ref{7recon}, right).  In these large  hot discs, the tidal arms
dominate the emission  lines and are  easily visible both  in the line
profiles directly and in Doppler tomograms.

\begin{figure}

\psfig{figure=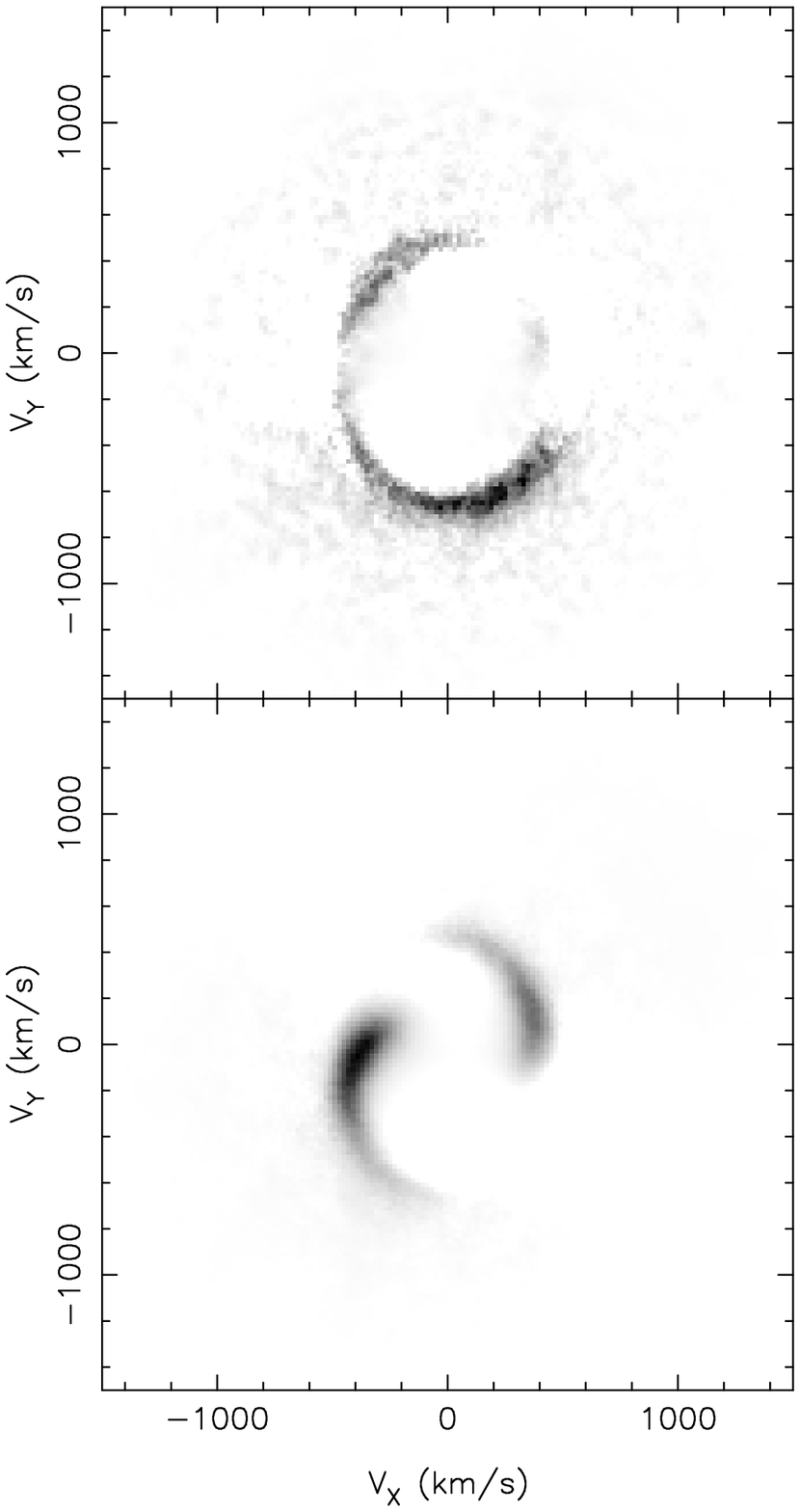,width=8cm}
\caption{The asymmetric parts of the reconstructed imaged for the two
models. In the high M case, only a slight asymmetry is produced in the
top left and lower right regions. In the low M case, the image is 
dominated by the two armed spiral structure.}
\label{asym}
\end{figure}

To  compare the   two models,  we show  the   asymmetric part  of  the
reconstructed tomograms in Fig.  \ref{asym}.    These are obtained  by
subtracting the median of the image  at each velocity, measured from the
white dwarf. The low Mach  number case (bottom) shows prominent spiral
arms,  whereas for  model I, only  a slight  asymmetry is  observed.

\subsection{Instrumental requirements}

We   reconstructed disc images  for  various  signal  to noise levels,
amounts of shear broadening, orbital inclinations and number of binary
phases using Model II data.   Based on these reconstructions, we found
a signal to  noise of  $\sim  15$ in  the line  flux and  a two  pixel
velocity   resolution of $\sim$80 km    s$^{-1}$  to be sufficient  to
convincingly   reproduce the two  armed  pattern  on the  disc. As  we
mentioned earlier,  to avoid  artifacts  $\sim$ 50  orbital phases are
desired.
The instrumental requirements for reconstructing such open spirals are
therefor  fortunately relatively modest and  suggest a search for such
structure in CV discs is feasible  and will provide strong constraints
on the presence of spiral waves in these discs.

\begin{figure*}
\psfig{figure=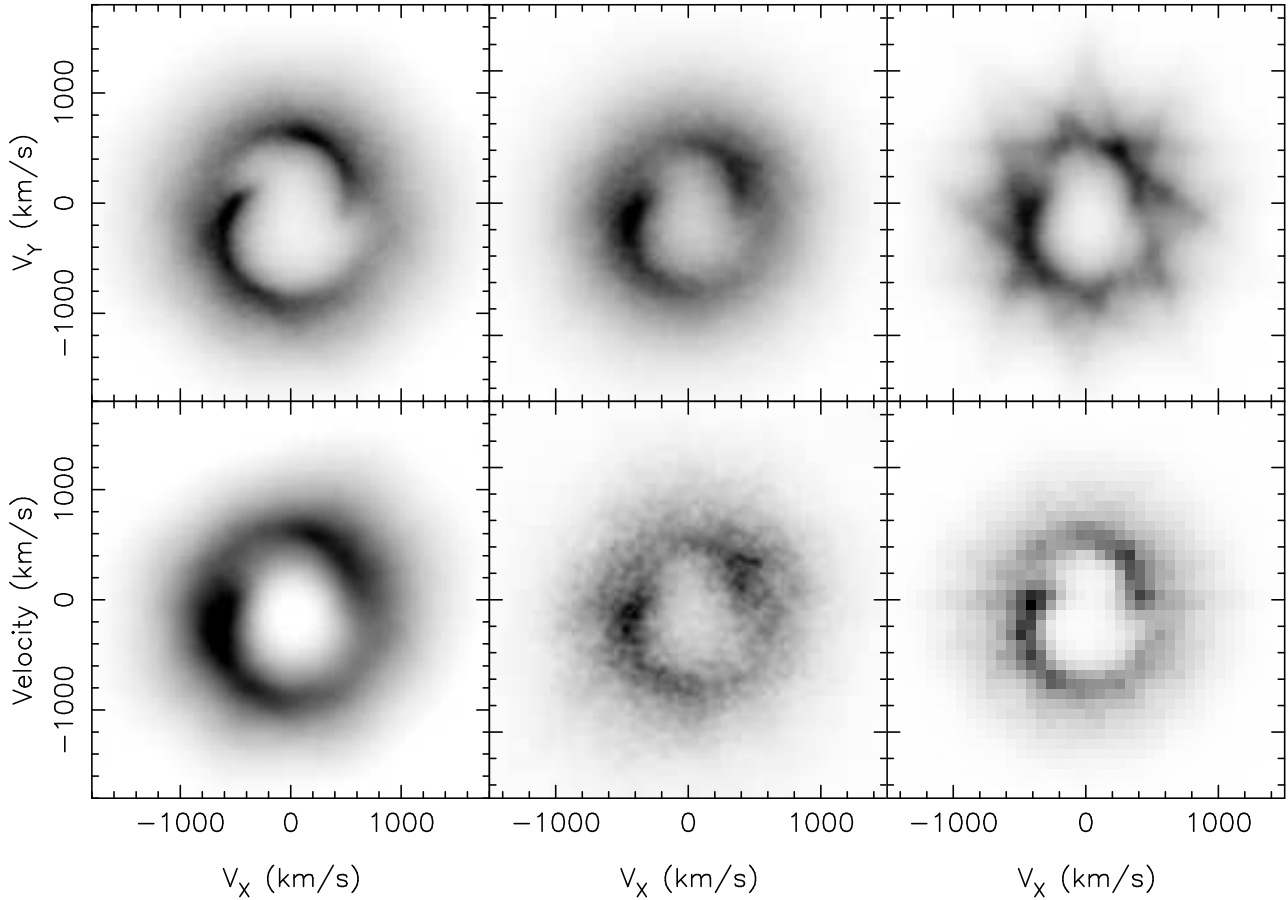,angle=-90,width=17cm}

\caption{Some example reconstructions for  Model II data.  Left  is 
the case of
high inclination (80$^{\circ}$) without (top)  and with maximal  shear
broadening.  Middle a reconstruction of a  strong emission line (twice
the continuum) with a continuum  signal to noise  of 30 (top) compared
to a weak line (15\% above continuum) at the same S/N. Top right shows
artifacts introduced by poor phase sampling, in this case 10 bins even
at high signal to noise  (50). Bottom right panel  is finally the case
of insufficient resolution (150 km  s$^{-1}$) to reconstruct the  disc
structure.}
\label{norecon}
\end{figure*}

Figure
\ref{norecon} illustrates some of the reconstructions obtained.
The left two images illustrate the effect of shear broadening for high
inclination cases.  The disc image  is similarly broadened in the case
of strong shear  broadening but still carries  the  two armed pattern.
Middle panels  illustrate the effect  of a strong versus weak emission
line.  This basically lowers   the signal to  noise  of the data   and
affects the disc image in the same way.  Finally the right panels show
the strong artifacts due  to poor phase sampling (top)  or the lack of
image structure due to poor resolution.

\subsection{Other emission line sources}

Our reconstructed images are those  of the accretion disc contribution
only.  The irradiated companion star, the  gas stream and the hot spot
are  well known other contributors to  the emission line flux.  In the
case of  the companion star, its  contribution is  confined to a Roche
lobe  shape  centered at  the radial   velocity  of the  companion, at
velocities where no disc emission is expected.  More demanding are the
stream-disc  interactions,  which can  contribute to a  wider range of
velocities overlapping the accretion  disc, though the compact  impact
region is usually well defined.   Complicated are disc outflows, which
can  be responsible for the  single peaked emission  lines observed in
some systems. These issues  have to be born  in mind
when interpreting emission line data sets (e.g. Steeghs, Horne, Marsh
\& Donati 1996).

\section{Discussion}
\label{discussion}

The Doppler  images and line profiles  presented in this paper, confirm
that   the observed disc  structure in  tomograms  of IP Pegasi can be
reproduced by tidally excited spiral arms in the disc.  The presence
of a dominating two armed structure in early outburst, when the disc is hot
and large, compared to  a fairly symmetric  disc in quiescence matches
extremely well with    the model  predictions  presented here.    
%
%
Armitage \& Murray (1998) reach a similar conclusion based on 3--D SPH
simulations of the disc flow in IP  Pegasi.  A spiral density pattern,
matching the kinematics of  the observations, is formed when  $\alpha$
is increased and the disc  expands.  However, these  arms appear to be
transient in their simulations on a viscous time scale.  In the models
of  Stehle (1999) the  spiral  pattern  is  stable in  the co-rotating
frame,  though its properties will  change as  the outburst progresses
and especially   when the cooling wave   starts to turn  the disc into
quiescence   again.  Such time dependent behaviour    can be tested by
Doppler mapping of the disc as the outburst progresses.
Godon,  Livio \&  Lubow  (1998) presented  2--D  disc simulations
using Fourier--Chebyshev spectral methods. They 
attempt to reproduce the observations of IP Pegasi. 
They find, however, that the spiral pattern resembles the observations
only for very high temperatures.
Unfortunately their $\alpha$ values were rather low ($\alpha =0.1$ at
most) and thus viscous spreading was less efficient compared to our
simulations.
As a consequence their discs where smaller by up to $\sim 50$\% 
compared to our model II simulation. 
Indeed, our model I simulation shows that
in the  case of low $\alpha$ values
little evidence  of spiral
arms in the emission lines is expected. However, 
with efficient viscous spreading the disc becomes large enough to 
result in a spiral pattern comparable to observed Doppler tomograms.

The behaviour of  the emission lines  in the presence of strong spiral
waves is very different   from   axisymmetric or elliptically     distorted
accretion discs without spiral shock waves.   In axisymmetric discs we
expect  a constant  separation between  the  peaks, and  in elliptical
discs   a smooth, sinusoidal    variation with orbital phase,  however
without sudden jumps  in the peak separation or the presence of multiple peaks
arising form successive spiral arms such as in Fig. \ref{outburst}.
Even  though spiral arms  in  close binary  accretion discs have  been
predicted more than 10 years ago (Sawada et al. 1986), and 
Doppler   tomography provides  a  way  of  resolving  the dynamics  of
accretion  discs,   a clear detection  of a   spiral  pattern was only
achieved  in a recent outburst of  IP Pegasi. Although confirmed since
in this particular object (Harlaftis et al. 1999), detection of such a
pattern in other systems is crucial.

We showed, however, that in small, cool discs, the observational 
signatures left by  tightly wound spiral  waves are  small.   
Indeed,  as Figure  \ref{5recon}  shows,   such discs, will    closely
resemble axisymmetric discs. 
The data on IP Pegasi and our simulations suggest that the increase in
disc  size and temperature during dwarf  novae outbursts  results in a
prominent  spiral  pattern in  the disc that   can  be recovered using
tomography.  The  number of Doppler  images of dwarf novae in outburst
is unfortunately small.
Marsh \&  Horne  (1990) provide Doppler  images  of IP Pegasi in  both
quiescence and  outburst. In  particular their  outburst  map  of HeII
($\lambda$4686)  four days into outburst show  a two armed enhancement
in the  disc  at the   correct  positions  in   the tomogram.    Their
resolution was  unfortunately rather   low ($\sim$130  km   s$^{-1}$).
Piche \& Szkody (1989) note that their outburst  data of IP Peg is not
compatible with a  symmetric  Keplerian  disc,  but no disc  image  is
available. Similarly,  Hessman (1989)   reports measurements  of  the
emission line peak  velocities of IP Pegasi  at the end of an outburst
which shows a phase dependence that suggests a strong asymmetry in the
outer disc persists well into the outburst, though is much weaker than
in the initial phases.
Outburst tomograms of SS Cyg (Steeghs  et al.  1996) show a
highly non-axisymmetric disc, again  with enhanced emission in the top
right and lower left areas.  On the other hand, tomography of the disc
in OY Car  in outburst (Harlaftis  \& Marsh 1996)  shows some evidence
for  the gas stream and  its impact on the disc,   but no clear spiral
disc structure.  Whether this is related to the  fact that OY Car is a
SU UMa type dwarf nova, will be explored in the future. 
Nova-like variables   provide permanent high   mass transfer accretion
discs, and we would therefore expect tidal structures to be important.
Unfortunately    the  emission lines from  these    systems are poorly
understood. The lines are single peaked  and contain S-wave components
of unknown   origin.   The exception could  be   V347  Pup,  the  only
eclipsing nova-like variable with  clear  double peaked emission  line
profiles originating in the disc (Still et al.   1998, Steeghs et al.,
in preparation).  The disc  appears to be  tidally distorted, and more
data may confirm the presence of spiral structure in this system.

Conclusive remarks on the presence of  spiral waves in accretion discs
are therefor difficult to make with the sparse amount of outburst data
available at this time. 
We have discussed the properties of CV discs since these are still the
best   laboratories to study  the   properties of  thin, $\alpha$ type
accretion discs where most heating is due to  viscous dissipation.  In
particular,  high    quality spectroscopic  studies of     dwarf novae
throughout their outburst  cycles are invaluable in understanding  the
physics   of thin accretion discs.   It  is important  to confirm with
further observations that  the spiral  density distribution  in  close
binary accretion discs is a common  phenomenon and during which source
states they can be observed. 

The   significance of tidal  waves   in the   discs of X-ray  binaries
(Blondin \& Owen  1997, Owen \&  Blondin 1997, Lanzafame  \& Belvedere
1998) is also considerable, since  the Mach numbers  in these discs are
much  lower and the relevance  of density waves  to  the total angular
momentum budget of the disc is  therefore greater compared  to CVs.  In
due course this should be accessible to similar observational tests.

\section{Summary}
\label{outlook}

We studied the emission line profiles of  close binary accretion discs
in cases where the disc pattern is dominated by prominent spiral shock
arms excited by the tidal forces of the secondary star.
As underlying  disc  models we used  the grid  of full hydrodynamic CV
disc calculations of Stehle (1999). We model  the emission lines using
the theory  of Horne (1995), including  the effect of shear
broadening for saturated lines.
Doppler  tomograms     and trailing  spectra where     constructed and
instrumental effects were    included to assess  the  data  quality
required to reconstruct spiral structure in CV discs.

We found that for high Mach number discs ($M=v_{\phi}/c_{\mathrm{s}}
\simeq $15 -- 30)
the  spiral shock  arms are too   tightly wound  to leave  significant
fingerprints in the emission line structure of these small discs.
As a result the peak separation of the double peaked emission line profile 
varies only by $\sim 8 $\% with binary phase.
We  therefore  do not  expect  to detect    strong spiral arms  in  
quiescent CV discs.

For low Mach number discs ($M \simeq $ 5 -- 20), i.e. for CV accretion
discs in   outburst, the line  emissivity is  dominated by  the spiral
shock arms. The two armed spiral pattern is identified by the presence
of converging peaks in the line profiles.
This signature in the trailing spectrum is expected to be sufficiently
prominent that a signal to noise of $\sim  15$ per pixel, a wavelength
resolution  of  $\sim 80$  km  s$^{-1}$ and a   time  resolution of 50
spectra per binary   orbit is sufficient to   verify  the presence  of
spiral   shock arms in      binary  accretion discs  through   Doppler
tomography of strong emission lines.

Our simulations successfully reproduce the emission line properties of
the disc structure observed in IP Pegasi.  The detection of strong two
armed spiral arms in outburst (Steeghs et al. 1997), but no clear disc
structure in  quiescence    (Marsh \&  Horne  1990),  is  convincingly
reproduced by tidal spiral waves in the accretion disc.

The prospect of a new way to probe the local physics of discs through the
dynamics of these tidal spirals demands more observational efforts on the
detailed properties of high-mass transfer accretion discs through indirect
imaging.

\section*{Acknowledgements}
We thank K.~Horne and A.~King for many stimulating discussions and for
improving the language  of the manuscript. Tom Marsh is  thanked for
the use of his {\sc DOPPLER} package.
R.S acknowledges a PPARC Rolling Grant for Theoretical Astrophysics 
to the Astronomy Group at Leicester.

\label{lastpage}
 
\end{document}